\documentclass[preprint]{aastex63}

\usepackage{hyperref}

\received{January 6, 2021}
\revised{February 23, 2021}
\accepted{February 24, 2021}
\submitjournal{ApJ}

\shorttitle{Hemispheric Sign Preference and Flaring Activity}
\shortauthors{Park et al.}

\graphicspath{{./}{figures/}}

\begin{document}

\title{Magnetic Helicity Flux across Solar Active Region Photospheres: II. Association of Hemispheric Sign Preference with Flaring Activity during Solar Cycle 24}

\correspondingauthor{Sung-Hong Park}
\email{shpark@isee.nagoya-u.ac.jp}

\author[0000-0001-9149-6547]{Sung-Hong Park}
\affiliation{Institute for Space-Earth Environmental Research, Nagoya University, Nagoya, Japan}

\author[0000-0003-0026-931X]{K. D. Leka}
\affiliation{Institute for Space-Earth Environmental Research, Nagoya University, Nagoya, Japan}
\affiliation{NorthWest Research Associates, Boulder, CO, USA}

\author[0000-0002-6814-6810]{Kanya Kusano}
\affiliation{Institute for Space-Earth Environmental Research, Nagoya University, Nagoya, Japan}

\begin{abstract}
In our earlier study of this series (Park et al. 2020, Paper I), we examined the hemispheric sign preference (HSP) of magnetic helicity flux $dH/dt$ across photospheric surfaces of 4802 samples of 1105 unique active regions (ARs) observed during solar cycle 24. Here, we investigate any association of the HSP, expressed as a degree of compliance, with flaring activity, analyzing the same set of $dH/dt$ estimates as used in Paper I. The AR samples under investigation are assigned to heliographic regions (HRs) defined in the Carrington longitude-latitude plane with a grid spacing of 45{\degr} in longitude and 15{\degr} in latitude. For AR samples in each of the defined HRs, we calculate the degree of HSP compliance and the average soft X-ray flare index. The strongest flaring activity is found to be in one distinctive HR with an extremely low HSP compliance of 41\% as compared to the mean and standard deviation of 62\% and 7\%, respectively, over all HRs. This sole HR shows an anti-HSP (i.e., $<$\,50\%) and includes the highly flare-productive AR NOAA 12673, however this AR is not uniquely responsible for the HR's low HSP. We also find that all HRs with the highest flaring activity are located in the southern hemisphere, and they tend to have lower degrees of HSP compliance. These findings point to the presence of localized regions of the convection zone with enhanced turbulence, imparting a greater magnetic complexity and a higher flaring rate to some rising magnetic flux tubes.
\end{abstract}

\keywords{methods: data analysis --- methods: observational --- Sun: activity --- Sun: flares --- Sun: magnetic fields --- Sun: photosphere}

\section{Introduction} \label{sec:intro}
Solar flares are known as the sudden and rapid release of magnetic energy stored in the solar atmosphere, producing electromagnetic radiation from radio waves to gamma rays. The vast majority of large flares occur in active regions (ARs) where intense magnetic fields manifest as dark sunspots in the photosphere. Many studies have been conducted to find out significant statistical differences of magnetic field properties between flaring and flare-quiet ARs \citep[e.g., see][and references therein]{1987SoPh..113..267Z,1989SoPh..121..197L,1994SoPh..149..105S,2007ApJ...671..955L,2007ApJ...656.1173L,2007ApJ...661L.109G,2008ApJ...686.1397P,2019LRSP...16....3T}. It is now generally accepted that flares are more likely to occur in ARs with magnetic fields exhibiting (1) more complex morphologies (e.g., $\delta$-sunspots), (2) higher degrees of magnetic non-potentiality in a wide variety of forms (e.g., magnetic twist or shear), and (3) more drastic changes in the photosphere near magnetic polarity inversion lines (e.g., magnetic flux emergence or cancellation). Even though these comparison studies have broadly advanced our understanding of flare-productive ARs, contemporary usage in the context of forecasting solar flares demonstrates that there are still fundamental aspects yet to be understood \citep[e.g.,][]{2016ApJ...829...89B,2017ApJ...835..156N,2019ApJS..243...36L,2020Sci...369..587K}. For example, it is still not clear whether one or many different critical conditions exist for triggering flares and driving eruptive instabilities, not to mention our continuing quest to understand the physical processes of flare energy build-up and release in the corona.

Magnetic helicity has received substantial attention over the past few decades, mainly due to its practical use in quantitatively measuring twists, kinks, and inter-linkages of magnetic field lines within a volume enclosed by a boundary surface $S$ \citep[e.g.,][]{1984JFM...147..133B,2006JPhA...39.8321B,2007ApJ...668..571L,2009AdSpR..43.1013D}. Magnetic helicity is gauge invariant only in the case of a closed volume satisfying that the normal component $B_{n}$ of the magnetic field vanishes at every point on $S$. In an open volume (i.e., $B_{n}$\,$\neq$\,0 at any point on $S$), on the other hand, a gauge-invariant form of the so-called \textit{relative} magnetic helicity was introduced with the choice of a reference magnetic field having the same $B_{n}$ at $S$. The relative magnetic helicity is defined as subtracting magnetic helicity of the reference field from the total \citep[refer to][]{1984JFM...147..133B,1985CoPPC..9..111}. For notational simplicity, hereafter, we refer to the relative magnetic helicity as magnetic helicity.

Since magnetic helicity flux $dH/dt$ across photospheric surfaces of ARs was first estimated from observations of the solar magnetic field \citep[e.g,][]{2001ApJ...560L..95C}, there have been many attempts to investigate $dH/dt$ in the context of understanding the energy build-up process and initiation mechanism of flares. For example, \citet{2010ApJ...718...43P} examined the temporal variation of $dH/dt$ over 24 hr for each of 378 unique ARs observed during solar cycle 23. They found a tendency that ARs with larger values of $|dH/dt|$ are more flare-productive. Meanwhile, temporal variations of $dH/dt$ over different timescales of a few hours to days before flare occurrence have been also investigated; consequently, some flare-associated variations of $dH/dt$ were found, including a large increase or sign reversal of $dH/dt$ \citep[e.g.,][]{2008ApJ...686.1397P,2014ApJ...794..118R,2017A&A...597A.104V}. It should be noted, however, that all these studies were carried out based on estimates of $dH/dt$ for a set of individual ARs. However, so far, no studies have tackled the question of whether the level of flaring activity has any association with the distribution of $dH/dt$ for a group of ARs located at the same region arbitrarily defined in the Carrington longitude-latitude heliographic plane.

In our earlier study of this series \citep[][Paper I]{2020ApJ...904....6P}, we explored the well-known hemispheric sign preference (hereafter referred to as HSP) of magnetic helicity: i.e., a dominance of negative (left-handed) helicity in the northern hemisphere and positive (right-handed) helicity in the southern hemisphere, independent of the solar cycle. More specifically, we examined the HSP, expressed as a degree of compliance that ARs follow the expected preference, analyzing $dH/dt$ across photospheric surfaces of ARs observed from 2010 to 2017 of solar cycle 24 by the Helioseismic and Magnetic Imager \citep[HMI,][]{2012SoPh..275..207S} on board the Solar Dynamics Observatory \citep[SDO;][]{2012SoPh..275....3P}. In Paper I, the HSP of $dH/dt$ was found to be stronger in the case of ARs that (1) appear at higher latitudes during the rising phase of the solar cycle; (2) have larger values of $|dH/dt|$, the total unsigned magnetic flux $\Phi$, and the average plasma-flow speed $<\!\!|v|\!\!>$\,$=$\,$<\!\!\sqrt{v_{x}^2+v_{y}^2+v_{z}^2}\!\!>$ through the given AR surface, where $v_{x}$, $v_{y}$ and $v_{z}$ are all three components of the photospheric magnetized plasma velocity derived from HMI vector magnetograms with the Differential Affine Velocity Estimator for Vector Magnetograms \citep[DAVE4VM;][]{2008ApJ...683.1134S}. These observed HSP dependencies suggest that the Coriolis force acting on a rising and expanding flux tube in the convection zone may play an important role in enhancing the HSP. Moreover, the HSP for ARs at higher latitudes may be strengthened by the differential rotation on the solar surface as well as the tachocline $\alpha$-effect of a flux-transport dynamo. With the same set of $dH/dt$ estimates as used in Paper I, here we study whether there is any relation between the HSP of $dH/dt$ and flaring activity. The $dH/dt$ data set is described in Section~\ref{subsec:dhdt_data_set}, the association of flares with their source ARs is presented in Section~\ref{subsec:flare_assignment}, and analysis results in Section~\ref{sec:results}. Finally, in Section~\ref{sec:discussion}, we summarize and discuss our main findings.

\section{Data Analysis} \label{sec:data_analysis}
\subsection{Description of the Data Set} \label{subsec:dhdt_data_set}
In this study, we analyze the same data set as in Paper I, which contains estimates of $dH/dt$, $\Phi$ and $<\!\!|v|\!\!>$ for all 4802 samples of 1105 unique NOAA-numbered ARs. The AR samples comprise pairs of vector magnetograms acquired from HMI AR Patches \citep[HARPs;][as recorded in the {\tt hmi.M\_720s} series]{2014SoPh..289.3483H}, and observed daily at 00:36 and 00:48 TAI over an 8 yr period from 2010 May 1 to 2017 December 3, satisfying the following criteria: (1) the longitudinal boundaries of the given HARP are located within $\pm$60$^\circ$ from the central meridian of the solar disk, and (2) the HARP contains only a single NOAA-numbered AR with at least one sunspot visible in white-light. Details of HMI vector magnetic field data and methods used to obtain $dH/dt$ estimates as well as their uncertainties can be found in Paper I and references therein. We note that the results in Section~\ref{sec:results} are only minimally affected by the uncertainties in $dH/dt$, because the uncertainties were found to be insufficient to cause the misidentification of the sign of $dH/dt$, as discussed in detail in Paper I.

For all of the AR samples, we find that 28\%, 57\%, and 15\% are $\alpha$-class, $\beta$-class, and the other complex-class ARs, respectively, according to their Mount Wilson (or Hale) magnetic classifications for sunspot groups. It is also found that 63\% of 2530 AR samples in the northern hemisphere and 65\% of 2272 samples in the southern hemisphere comply with the HSP of $dH/dt$. These observed degrees of HSP compliance are within the range reported in previous studies \citep[e.g.,][]{1998ApJ...507..417L,2001ApJ...549L.261P,2005PASJ...57..481H,2006ApJ...646L..85Z,2014ApJ...783L...1L}. We refer the reader to Paper I for the HSP dependencies identified with respect to various properties of ARs as well as some relevant physical mechanisms for adherence to the observed HSP.

\subsection{Assignment of Flares to the Active Region Samples} \label{subsec:flare_assignment}
The NOAA/Space Weather Prediction Center (SWPC) provides historical flare event list data (\url{ftp://ftp.swpc.noaa.gov/pub/warehouse}), obtained by the Geostationary Operational Environmental Satellites (GOES), along with a network of ground-based solar observatories. The SWPC flare event list contains relevant information, such as flare start times, magnitudes, source regions and locations, in order to find GOES soft X-ray flares that occurred in a given AR within an interval $\tau$\,$=$\,24 hr following the AR observation time (00:36\,TAI). For each of our AR samples, the flare assignment is done basically searching for flares within $\tau$, of which source regions are assigned with the same NOAA region number as the AR sample. In the SWPC list, there are however some flares for which no information is available on their source regions (i.e., no NOAA region number is given), but their heliographic locations (i.e., longitudes and latitudes) are available. Among such flares within $\tau$, those which are located within the derotated HARP field of view (FOV) of the given AR sample at the flare start times are assigned to that AR sample.

To quantify flaring activity of each AR sample under consideration, we define the 24 hr flare index (hereafter shortly indicated as $F_{idx}$) as
\begin{equation}
F_{idx} = 100 \times S^{X} + 10 \times S^{M} + 1 \times S^{C} + 0.1 \times S^{B},
\label{eq:flare_index}
\end{equation}
where $S^{j} = \sum_{i=1}^{N_j} \mathcal{M}_{i}^{j}$. Here $N_j$ is the total number of $j$-class flares assigned to the given AR sample during $\tau$, and $\mathcal{M}_{i}^{j}$ is the magnitude (i.e., digit multipliers from 1.0 to 9.9) of the $j$-class flares. Simply put, $F_{idx}$ refers to the sum of GOES soft X-ray peak fluxes of all flares assigned to the given AR during $\tau$, and it is often considered to measure flare productivity of an AR for a target interval \citep[e.g.,][]{2005ApJ...629.1141A,2010ApJ...718...43P,2018SoPh..293..159L}.

\subsection{Calculation of the HSP for Defined HRs} \label{subsec:heliographic_regions}
We define heliographic regions (HRs) in the Carrington longitude-latitude plane, each of which has the longitudinal and latitudinal extents of 45{\degr} and 15{\degr}, respectively. The AR samples are then assigned to the defined HRs, based on where the center coordinates of the AR samples are located. In the case that the AR sample's HARP FOV spans two or more HRs, the assignment is to that which includes the largest fraction of the HARP. For each HR, the degree of HSP compliance is then calculated as the fraction of AR samples that have negative/positive values of $dH/dt$ if the given HR is located in the northern/southern hemisphere. 

\section{Results} \label{sec:results}
\subsection{HSP versus Anti-HSP Active Region Properties} \label{subsec:hsp_vs_anti_hsp}
\begin{figure}[t!]
\centering
\includegraphics[width=0.84\textwidth]{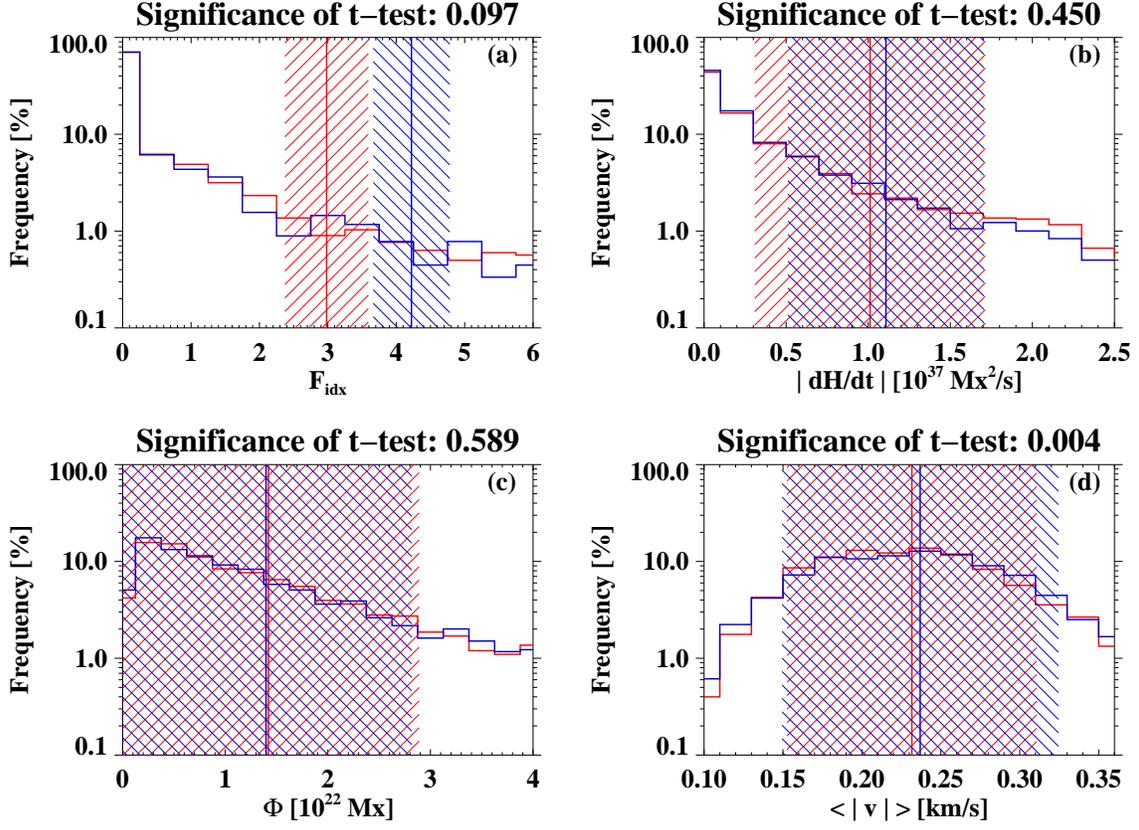}
\caption{Frequency distributions of 3007 HSP AR samples (red curve) and 1795 anti-HSP AR samples (blue curve) considered in this study, with respect to (a) the flare index $F_{idx}$, (b) the absolute value of $dH/dt$, (c) the total unsigned magnetic flux $\Phi$, and (d) the AR samples' average plasma-flow speed $<\!\!|v|\!\!>$. In each panel, the mean (vertical line) and mean\,$\pm$\,IQR (tilted lines) are marked for the HSP and anti-HSP samples, respectively, where IQR is the interquartile range defined as the 75th percentile minus 25th percentile.}
\label{fig:hsp_vs_anti_hsp}
\end{figure}

With the given set of our $dH/dt$ estimates, we first examine whether the level of flaring activity is different between two separate groups of AR samples: i.e., 3007 HSP AR samples with $dH/dt$ estimates following the HSP versus 1795 anti-HSP AR samples with $dH/dt$ estimates against the HSP. In panel (a) of Figure~\ref{fig:hsp_vs_anti_hsp}, the relative frequency distribution of $F_{idx}$ values is shown for the HSP samples (red curve) and anti-HSP samples (blue curve), respectively, with the mean (vertical line) and mean\,$\pm$\,IQR (tilted lines), where IQR is the interquartile range defined as the 75th percentile minus 25th percentile. We find that the mean of $F_{idx}$ values for the anti-HSP AR samples is larger than that for the HSP samples. The difference between them is however not significant according to the two-sided $p$-value of 0.097 from the Student's t-test. In each case of the parameters $|dH/dt|$ (panel (b)) and $\Phi$ (panel (c)), the mean value for the HSP samples is found to be similar to that for the anti-HSP samples in the context that: (1) $p$-value is much larger than 0.1, and (2) the mean value for the HSP samples is well constrained within the range of the mean\,$\pm$\,0.2\,$\times$\,IQR for the anti-HSP samples, and vice versa. Meanwhile, as shown in panel (d) of Figure~\ref{fig:hsp_vs_anti_hsp}, the anti-HSP samples have larger values of $<\!\!|v|\!\!>$ generally compared to the HSP samples with the $p$-value of 0.004, but the difference between the mean values is within 0.1\,$\times$\,IQR for either the HSP samples or the anti-HSP samples.

Flares are known to occur more frequently in ARs that include complex magnetic fields, such as $\delta$-sunspots \citep[e.g.,][]{1987SoPh..113..267Z,1989SoPh..121..197L,1994SoPh..149..105S,2000ApJ...540..583S,2012SoPh..281..639L}. In this respect, we investigate what percentage of the HSP and anti-HSP AR samples, respectively, belong to $\delta$-class, and whether there is a notable difference of the $\delta$-class percentages between the two AR groups. We find 4.9$\pm$1.8\% of the HSP AR samples are found to be categorized as having $\delta$-sunspots, while for the anti-HSP AR samples it is 5.0$\pm$2.3\%. The error in each $\delta$-class percentage is estimated by the Poisson uncertainty of $1/\sqrt{N}$, where $N$ is the total number of the given samples. We find that the $\delta$-class percentage shows no significant difference between the two groups, even though the mean of $F_{idx}$ is slightly larger for the anti-HSP samples compared to the HSP samples.

\subsection{Association of the HSP with Flaring Activity in Heliographic Regions} \label{subsec:hsp_vs_anti_hsp}
\begin{figure}[t!]
\centering
\includegraphics[width=0.89\textwidth]{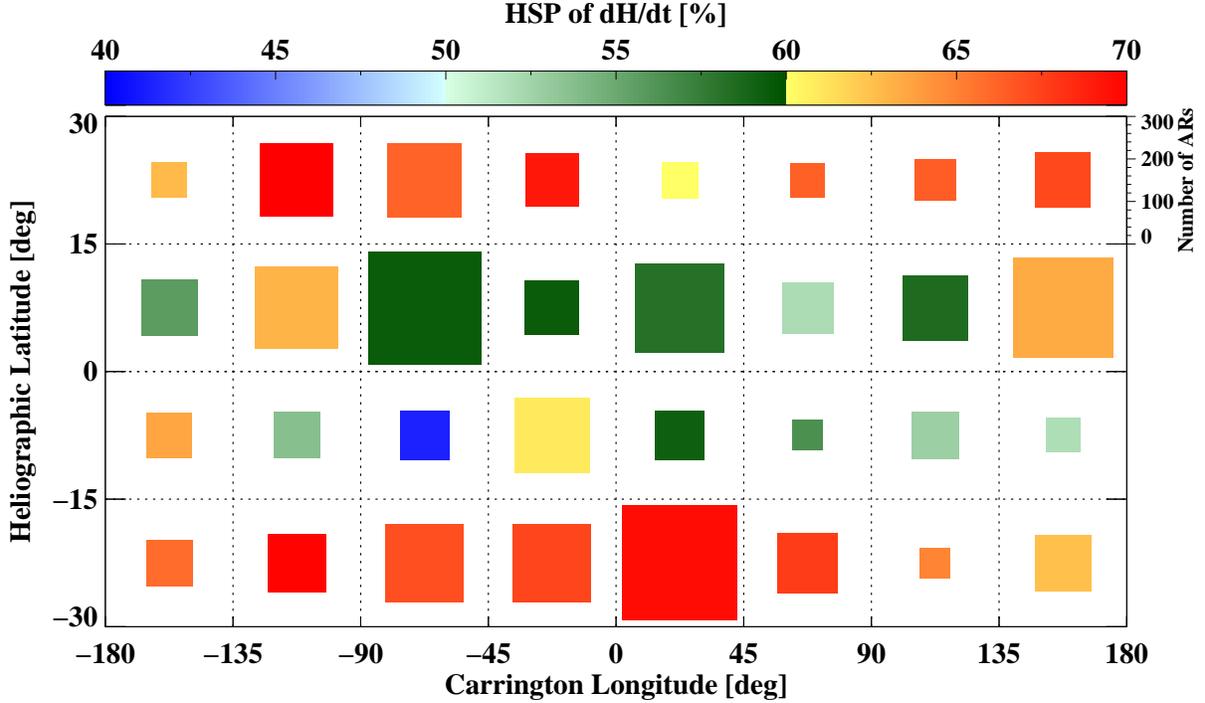}
\caption{Each heliographic region (HR) defined in this study is color-coded by the degree of HSP compliance of $dH/dt$. The side length of the squares represents the number of AR samples located at the defined HRs.}
\label{fig:diagram_hsp}
\end{figure}

Here, we aim to find any association of the HSP with flaring activity in the HRs defined in Section~\ref{subsec:heliographic_regions}. In Figure~\ref{fig:diagram_hsp}, each of the defined HRs is color-coded by the degree of HSP compliance for the subset of our AR samples contained therein. The number of AR samples located within a given HR is indicated by the side length of the color-coded square in that HR. In this HSP diagram, higher degrees of HSP compliance (i.e., $\sim$60\,--\,70\%) are found in the regions at higher latitudes in the ranges [$-$30{\degr},\,$-$15{\degr}] and [15{\degr},\,30{\degr}], compared to lower degrees of HSP compliance (i.e., $\sim$40\,--\,60\% in most cases) in the regions at lower latitudes [$-$15{\degr},\,0{\degr}] and [0{\degr},\,15{\degr}]. Such latitudinal dependence of the HSP was reported in Paper I. More interestingly, we find a notable HR at [$-$90{\degr},\,$-$45{\degr}] in Carrington longitude and [$-$15{\degr},\,0{\degr}] in latitude, exhibiting an extremely low HSP compliance of 41\% as compared to the mean and standard deviation of 62\% and 7\%, respectively, for all HRs. It should be also noted that this HR is the only one that definitively shows an anti-HSP (i.e., less than 50\% compliance).

\begin{figure}[h!]
\centering
\includegraphics[width=0.89\textwidth]{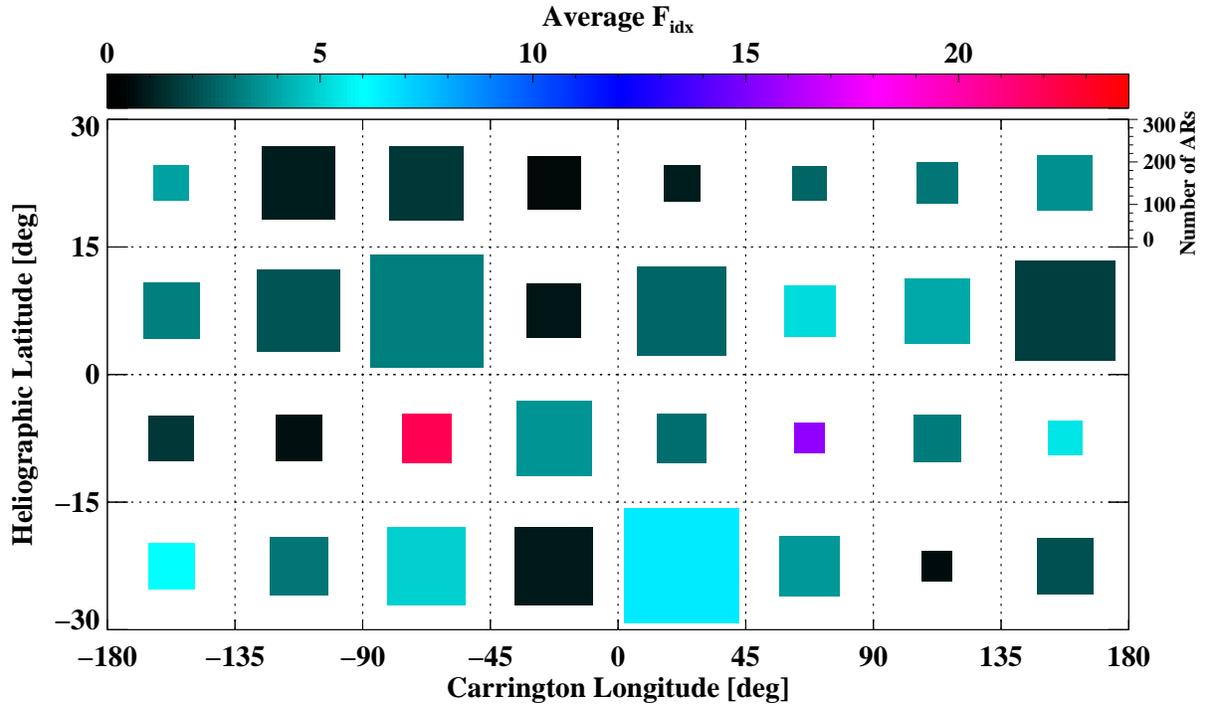}
\caption{Same as Figure~\ref{fig:diagram_hsp}, but showing the average value of the 24 hr flare index $F_{idx}$ for AR samples assigned to each of the defined HRs.}
\label{fig:diagram_fidx}
\end{figure}

\begin{figure}[h!]
\centering
\includegraphics[width=0.89\textwidth]{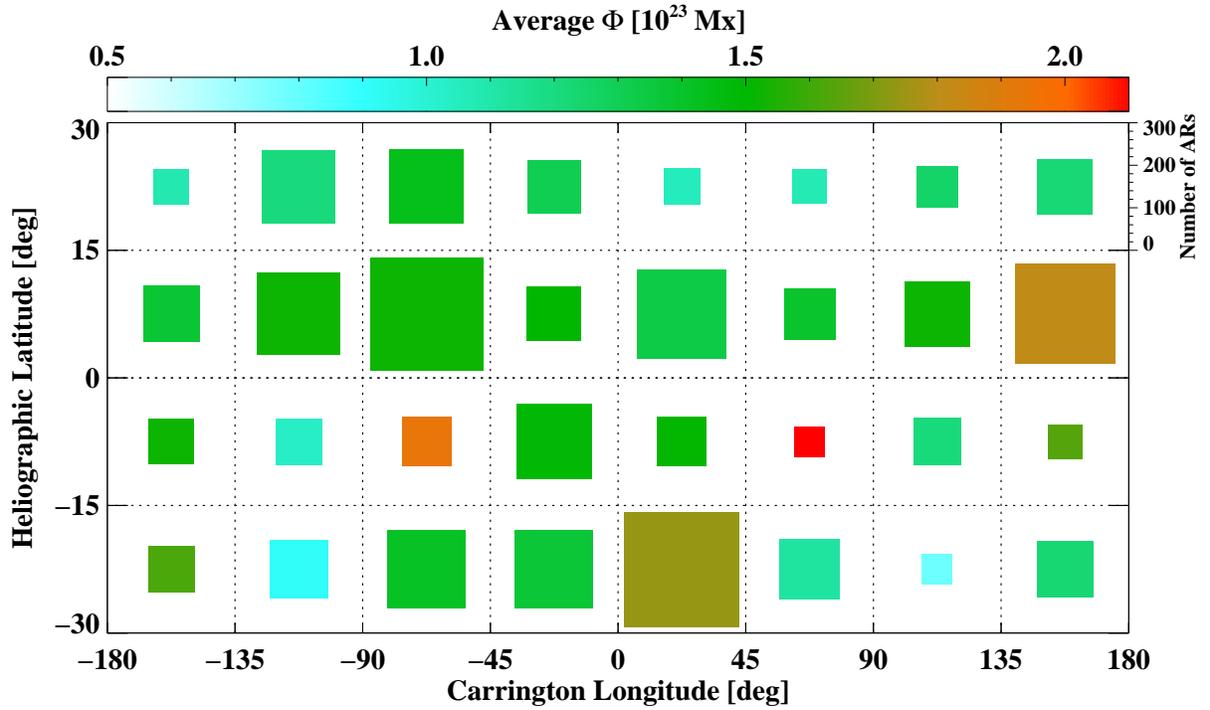}
\caption{Same as Figure~\ref{fig:diagram_hsp}, but showing the average value of the total unsigned magnetic flux $\Phi$ for AR samples assigned to each of the defined HRs.}
\label{fig:diagram_tuflux}
\end{figure}

Figure~\ref{fig:diagram_fidx} shows the average $F_{idx}$ values for the same subset of AR samples assigned to each of the HRs as in Figure~\ref{fig:diagram_hsp}. We find that the anti-HSP HR with the remarkably low HSP compliance shows the strongest flaring activity, having the largest value of the average $F_{idx}$\,$=$\,22. This is extremely large relative to the mean and standard deviation (i.e., 3.8 and 4.3, respectively) of the average $F_{idx}$ values for all HRs. This anti-HSP HR contains the highly flare-productive AR NOAA 12673. However, even when NOAA 12673 is excluded, this HR still shows an anti-HSP with the lowest HSP compliance of 44\% as well as a high level of flaring activity with the average $F_{idx}$\,$=$\,6 ranked in the top three. Another large flare-productive HR with the average $F_{idx}$\,$=$\,15, including the largest AR NOAA 12192 observed in cycle 24, is found at [45{\degr},\,90{\degr}] in Carrington longitude and [$-$15{\degr},\,0{\degr}] in latitude. This region has a low HSP compliance of 56\% that is smaller than the mean minus one standard deviation. In addition, as shown in Figure~\ref{fig:diagram_tuflux}, those two specific HRs have the average $\Phi$ values ranked in the top two (i.e., greater than 1.9\,$\times$\,10$^{22}$\,Mx). 

\begin{figure}[t!]
\centering
\includegraphics[width=1\textwidth]{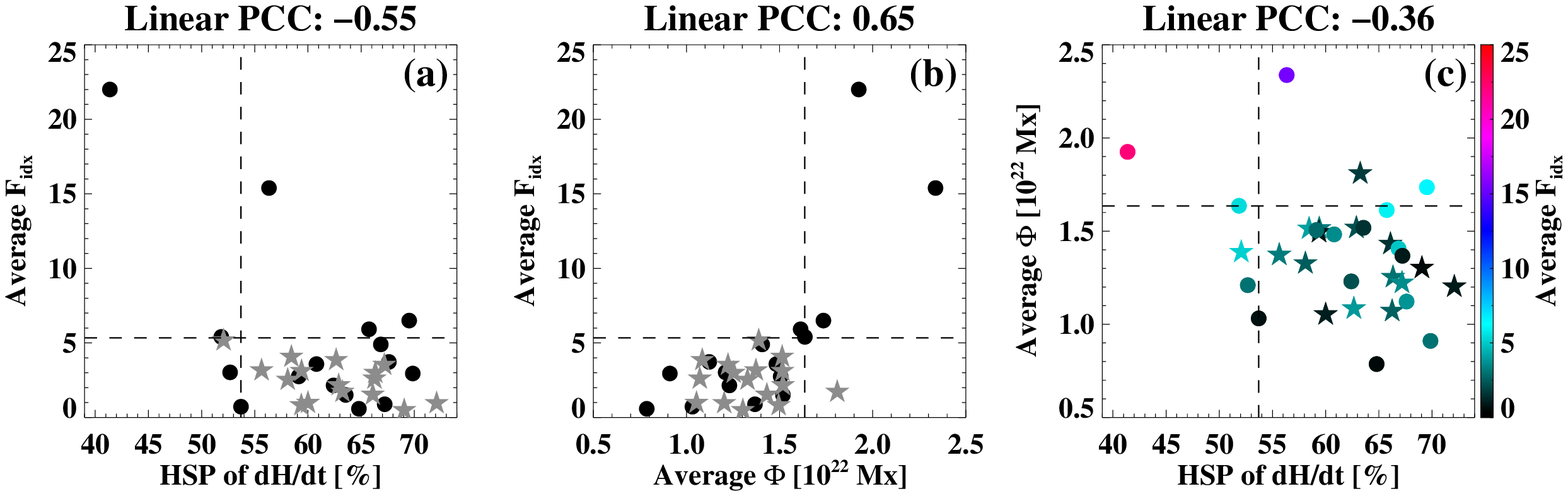}
\caption{Scatter plots of (a) the average $F_{idx}$ versus the HSP, (b) the average $F_{idx}$ versus the average $\Phi$, and (c) the average $\Phi$ versus the HSP for the defined HRs in the northern (stars) and southern (circles) hemispheres. In panel (c), the symbols are color-coded by the average $F_{idx}$. Data points of the five lowest HSP, the five largest average $\Phi$, or the five largest average $F_{idx}$, respectively, are separated from the others by the dashed lines in each panel.}
\label{fig:hsp_tuflux_fidx}
\end{figure}

Now we turn to the question of whether there are any general trends or strong correlations between the degree of HSP compliance, the average $F_{idx}$, and the average $\Phi$ derived from the AR samples contained in the defined HRs. As shown in the scatter plot of the average $F_{idx}$ versus the HSP (panel (a) of Figure~\ref{fig:hsp_tuflux_fidx}), we find a weak tendency that HRs with lower degrees of HSP compliance show larger values of the average $F_{idx}$, with the linear Pearson correlation coefficient (PCC) of $-$0.55. In the scatter plot, the anti-HSP region with the largest average $F_{idx}$ can be considered as an extreme case, which lies far away from both the vertical and horizontal dashes lines used to separate data points of the five lowest HSP and the five largest average $F_{idx}$, respectively. On the other hand, a positive correlation exists between the average $\Phi$ and the average $F_{idx}$ with the linear PCC\,$=$\,0.65 (refer to panel (b) of Figure~\ref{fig:hsp_tuflux_fidx}). Such correlation of $\Phi$ with flaring activity in ``individual'' ARs has been reported in many previous studies \citep[e.g.,][]{2003ApJ...595.1296L,2007ApJ...656.1173L,2010ApJ...718...43P,2017ApJ...843..104L,2018SoPh..293..159L}, but it is reported here for the first time on this larger spatial scale of the HRs over a much longer period of solar cycle 24. As shown in Figure~\ref{fig:hsp_tuflux_fidx}{(c)}, a negative, although weak, correlation appears between the HSP and the average $\Phi$ with the linear PCC\,$=$\,$-$0.36. In contrast, a weak trend was reported in Paper I that ARs with larger values of $\Phi$ show a higher HSP. These two contrasting trends of the HSP with the average $\Phi$ are mainly due to the different ways of dividing the same AR samples into smaller subsets. The AR subsets in this study were selected based on heliographic locations of the AR samples in the defined Carrington longitude-latitude plane, while those in Paper I as a function of $\Phi$ values. The two highest flare-productive HRs with the average $F_{idx}$\,$\ge$\,15 have the average $\Phi$ values ranked in the top two as well as lower degrees of HSP compliance (i.e., one with the lowest HSP and the other with the HSP in the bottom 20\%). Meanwhile, the two highly flare-productive regions can be considered as obvious outliers, compared to the positive trend of the HSP with respect to the average $\Phi$ in Paper I. This may indicate that the HSP for those regions is obscured by vigorous turbulent convective flows interacting with rising flux tubes therein. As mentioned earlier, the two X-class flaring ARs, NOAA 12673 and NOAA 12192, are located at the two HRs, respectively. Even excluding these two influential ARs, however, all of the trends as described in Figure~\ref{fig:hsp_tuflux_fidx} remain the same, although the correlations become less strong (i.e., the linear PCCs of $-$0.25, 0.51 and $-$0.27 for the cases in panels (a), (b) and (c) of Figure~\ref{fig:hsp_tuflux_fidx}).

\subsection{North-south Asymmetry of the HSP} \label{subsec:noth_south_asymmetry}
We explore the north-south hemispheric asymmetry of solar activity during solar cycle 24. The north-south asymmetry has been studied with respect to various solar activity indices for cycle 24, including the total number of flares for a given GOES class \citep[e.g.,][]{2017Ap.....60..387B,2019OAst...28..228J}, and sunspot areas \citep[e.g.,][]{2019ApJ...873..121L}. The scatter plots in Figure~\ref{fig:hsp_tuflux_fidx} are used to examine any notable asymmetry in the distributions of the HSP, the average $\Phi$ and the average $F_{idx}$ between the HRs in northern (stars) and southern (circles) hemispheres. For the HSP, as shown in Figure~\ref{fig:hsp_tuflux_fidx}{(a)}, we find that four out of the five HRs with the HSP ranked in the bottom five (i.e., bottom 15\%) are located in the southern hemisphere. A similar north-south asymmetry is found in the average $\Phi$ (see Figure~\ref{fig:hsp_tuflux_fidx}{(b)}), but for the HRs with the five largest average $\Phi$ (i.e., top 15\%). In the case of the average $F_{idx}$, all of the HRs in the top five are placed in the southern hemisphere. 

These observational findings lend support to the presence of the north-south asymmetric behavior of solar activity in cycle 24, as discussed above, in which the southern hemisphere is more active overall than the northern hemisphere. In the present analysis, this asymmetry is most clearly visible with the activity metrics used here (i.e., the HSP, the average $F_{idx}$, and the average $\Phi$) between the HRs lying within [$-$15{\degr},\,0{\degr}] versus those lying within [0{\degr},\,15{\degr}] (refer to Figures~\ref{fig:diagram_hsp}--\ref{fig:diagram_tuflux}). The observed asymmetry in favor of stronger activity in the southern hemisphere during cycle 24 agrees with other observations of more frequent flare events at C- or M-class by \citet{2019OAst...28..228J} as well as larger values of the yearly mean sunspot area by \citet{2019ApJ...873..121L}. In addition, the southern hemisphere is expected to be more active based on the previously reported periodicities of the north-south asymmetry longer than a few solar cycles (e.g., periodic behaviors of $\sim$4 and $\sim$12 cycles shown in \citet{2005A&A...431L...5B} and \citet{2002A&A...383..648L}, respectively). 

\section{Summary and Conclusions} \label{sec:discussion}
In this paper we have investigated the hemispheric sign preference (HSP) of magnetic helicity flux $dH/dt$ for 4802 samples of 1105 unique active regions (ARs), in the context of whether and how the HSP is associated with flaring activity of the AR samples observed from 2010 to 2017 of solar cycle 24. The AR samples were first categorized into two separate groups of 3007 HSP AR samples with $dH/dt$ estimates following the HSP and 1795 anti-HSP AR samples with $dH/dt$ estimates against the HSP. Comparing values of the 24 hr flare index $F_{idx}$ between the HSP and anti-HSP AR samples, we found that the mean $F_{idx}$ for the anti-HSP samples is larger than that for the HSP samples, albeit with minimal statistical significance. Next, heliographic regions (HRs) were defined in the Carrington longitude-latitude heliographic plane, each of which has the longitudinal and latitudinal extents of 45{\degr} and 15{\degr}, respectively. We then examined the relations between the degree of HSP compliance, the average total unsigned flux $\Phi$ and the average $F_{idx}$ for a subset of AR samples placed in each of the defined HRs. Our main findings can be summarized as follows:
\begin{enumerate}
  \item Among the HRs is a distinctive one with the strongest flaring activity that exhibits an extremely low HSP compliance of 41\% as compared to the mean and standard deviation of 62\% and 7\%, respectively, for all the regions. This HR includes the highly flare-productive AR NOAA 12673. Note that even if NOAA 12673 is excluded, the HR still remains an anti-HSP with the lowest HSP compliance of 44\%.
  \item There is a weak tendency for HRs with larger values of the average $F_{idx}$ to have lower degrees of HSP compliance with the the linear Pearson correlation coefficient (PCC) of $-$0.55, as well as larger values of the average $\Phi$ with the linear PCC of 0.65.
  \item All HRs whose average $F_{idx}$ values rank in the top 15\% are located in the southern hemisphere. Such hemispheric asymmetry is also found in favor of lower degrees of HSP compliance as well as larger values of the average $\Phi$ in the southern hemisphere.
\end{enumerate}

In summary, all these observational findings lend support to the general trend of stronger flaring activity in a given HR with a lower degree of HSP compliance during cycle 24. We now attempt to tackle the question of why there may exist such association of the HSP with flaring activity. \citet{2007ApJ...671..955L} compared $dH/dt$ estimates between 48 X-class flaring ARs and 345 ARs without X-class flares. Separating the components of $dH/dt$ into the one $dH/dt^\mathrm{rot}$ by the differential rotation on the solar surface and the rest $dH/dt^\mathrm{rest}$, they found that on average, the X-class flaring ARs have a larger ratio of $|dH/dt^\mathrm{rest}|$ to $|dH/dt^\mathrm{rot}|$, i.e., $\sim$7:1, than $\sim$5:1 for the ARs that did not produce any X-class flares. They also showed that $dH/dt^\mathrm{rest}$ can be more disorganized and of opposite sign compared to the HSP, so that the HSP is effectively obscured by $dH/dt^\mathrm{rest}$. 

These results suggest that the surface differential rotation is less likely to be a crucial mechanism to cause the observed X-class flares, but the other components contributing to $dH/dt$, such as an emerging twisted magnetic flux tube across the photosphere, may be more related to flare energy build-up and triggering mechanisms. In this respect, it is noteworthy to consider the ``$\Sigma$-effect'' \citep{1998ApJ...507..417L}, which acts on a buoyantly rising and expanding flux tube through the turbulent convection zone. Through numerical simulations of the $\Sigma$-effect on rising flux tubes, \citet{1998ApJ...507..417L} found that a lower degree of HSP compliance can be obtained mainly due to the interaction of a flux tube with increased magnitudes of turbulent velocities in the convection zone. Based on the $\Sigma$-effect simulations, we conjecture that the distinctive anti-HSP heliographic region found in the present study may have highly turbulent localized flows through its layers from the deep convection zone to the photosphere. Such localized, amplified turbulence in the convection zone may play a crucial role for rising flux tubes as their magnetic fields get more complex with larger magnetic non-potentiality, and eventually produce large flares. 

As shown in Figure~\ref{fig:hsp_tuflux_fidx}{(c)}, the inverse correlation between the HSP and the average $\Phi$ may also support the conjecture that the expected high-HSP for a rapidly expanding flux tube (with large magnetic flux) therein by the Coriolis force would be obscured by highly turbulent convective flows in such low-HSP HRs. Moreover, a number of numerical simulations showed that ARs with complex magnetic structures (e.g., $\gamma$- or $\delta$-class) are formed by multiple buoyantly emerging segments of a single subsurface flux tube \citep[e.g.,][]{2014SoPh..289.3351T,2015ApJ...806...79F,2017ApJ...850...39T}, confirming prior observational studies \citep[e.g.,][]{1996ApJ...462..547L,2013ApJ...764L...3C}. Such multiple buoyant segments of a single subsurface flux tube may be produced by intense turbulence of localized regions in the convection zone. Meanwhile, examining Carrington longitudes of recurrent magnetic flux emergence for several solar rotations (called ``long-lived activity complexes''), \citet{2015ApJ...798...20K} revealed that each such activity complex typically has a mixture of positive and negative current helicity over its lifetime. In these activity complexes, newly emerging magnetic flux interacts with pre-existing flux of prior ARs, which may cause an increase in magnetic field complexity and lead to flares at an enhanced rate.

The observational findings in this study reveal that there are regional differences of HSP compliance in relation to flaring activity during solar cycle 24. Similarly, localized enhancements in various types of solar activity over long-term intervals of a few years to several solar cycles were reported in earlier studies: e.g., active longitudes of sunspots \citep{1982SoPh...76..155B,1983ApJ...265.1056G,2003A&A...405.1121B,2017ApJ...835...62M}, preferred longitudinal domains of flares \citep[also known as hot spots;][]{1988ApJ...328..860B,1990ApJ...364L..17B,2003ApJ...585.1114B,2016ApJ...818..127G}, coronal mass ejections \citep{2017ApJ...838...18G} and kinetic helicity \citep{2019ApJ...887..192K}. Future studies on the solar dynamo as it operates in the convection zone on the global scale should explore how such localized variations can be generated, in particular, with respect to the observed non-homogeneity of HSP compliance over the HRs when a solar cycle is considered. We also expect that the results found in this study can be examined in detail with numerical simulations of magnetic flux emergence through the layers from the deep convection zone to the photosphere \citep[e.g.,][]{2019ApJ...886L..21T,2020MNRAS.498.2925H}. This will enlighten our understanding of ARs which may be born to produce flares even before their emergence on the surface, but at the time of their earlier formation in the shape of a twisted magnetic flux tube in the deep convection zone or during the rise of the flux tube through the convection zone interacting with both large-scale systematic flows (e.g., differential rotation, meridional circulation) and localized turbulent flows.

\acknowledgments
The authors would like to thank an anonymous referee for constructive comments, and the NorthWest Research Associates (NWRA)/Boulder team for the development and maintenance of the SDO remote-Storage Unit Management System(SUMS)/Data Record Management System(DRMS) system and analysis servers at NWRA. The data used in this work are courtesy of the NASA/SDO and the HMI science teams. This research has made extensive use of the NASA's Astrophysics Data System (ADS) as well as the computer system of the Center for Integrated Data Science (CIDAS), Institute for Space-Earth Environmental Research (ISEE), Nagoya University. This work was partially supported by MEXT/JSPS KAKENHI Grant No.~JP15H05814.
\vspace{5mm}
\facilities{SDO (HMI)}
\software{DAVE4VM \citep{2008ApJ...683.1134S}}


\begin{thebibliography}{}
\expandafter\ifx\csname natexlab\endcsname\relax\def\natexlab#1{#1}\fi
\providecommand{\url}[1]{\href{#1}{#1}}
\providecommand{\dodoi}[1]{doi:~\href{http://doi.org/#1}{\nolinkurl{#1}}}
\providecommand{\doeprint}[1]{\href{http://ascl.net/#1}{\nolinkurl{http://ascl.net/#1}}}
\providecommand{\doarXiv}[1]{\href{https://arxiv.org/abs/#1}{\nolinkurl{https://arxiv.org/abs/#1}}}

\bibitem[{{Abramenko}(2005)}]{2005ApJ...629.1141A}
{Abramenko}, V.~I. 2005, \apj, 629, 1141, \dodoi{10.1086/431732}

\bibitem[{{Bai}(1988)}]{1988ApJ...328..860B}
{Bai}, T. 1988, \apj, 328, 860, \dodoi{10.1086/166344}

\bibitem[{{Bai}(1990)}]{1990ApJ...364L..17B}
---. 1990, \apjl, 364, L17, \dodoi{10.1086/185864}

\bibitem[{{Bai}(2003)}]{2003ApJ...585.1114B}
---. 2003, \apj, 585, 1114, \dodoi{10.1086/346152}

\bibitem[{{Ballester} {et~al.}(2005){Ballester}, {Oliver}, \&
  {Carbonell}}]{2005A&A...431L...5B}
{Ballester}, J.~L., {Oliver}, R., \& {Carbonell}, M. 2005, \aap, 431, L5,
  \dodoi{10.1051/0004-6361:200400135}

\bibitem[{{Barnes} {et~al.}(2016){Barnes}, {Leka}, {Schrijver}, {Colak},
  {Qahwaji}, {Ashamari}, {Yuan}, {Zhang}, {McAteer}, {Bloomfield}, {Higgins},
  {Gallagher}, {Falconer}, {Georgoulis}, {Wheatland}, {Balch}, {Dunn}, \&
  {Wagner}}]{2016ApJ...829...89B}
{Barnes}, G., {Leka}, K.~D., {Schrijver}, C.~J., {et~al.} 2016, \apj, 829, 89,
  \dodoi{10.3847/0004-637X/829/2/89}

\bibitem[{{Berdyugina} \& {Usoskin}(2003)}]{2003A&A...405.1121B}
{Berdyugina}, S.~V., \& {Usoskin}, I.~G. 2003, \aap, 405, 1121,
  \dodoi{10.1051/0004-6361:20030748}

\bibitem[{{Berger} \& {Field}(1984)}]{1984JFM...147..133B}
{Berger}, M.~A., \& {Field}, G.~B. 1984, Journal of Fluid Mechanics, 147, 133,
  \dodoi{10.1017/S0022112084002019}

\bibitem[{{Berger} \& {Prior}(2006)}]{2006JPhA...39.8321B}
{Berger}, M.~A., \& {Prior}, C. 2006, Journal of Physics A Mathematical
  General, 39, 8321, \dodoi{10.1088/0305-4470/39/26/005}

\bibitem[{{Bogart}(1982)}]{1982SoPh...76..155B}
{Bogart}, R.~S. 1982, \solphys, 76, 155, \dodoi{10.1007/BF00214137}

\bibitem[{{Bruevich} \& {Yakunina}(2017)}]{2017Ap.....60..387B}
{Bruevich}, E.~A., \& {Yakunina}, G.~V. 2017, Astrophysics, 60, 387,
  \dodoi{10.1007/s10511-017-9492-7}

\bibitem[{{Chae}(2001)}]{2001ApJ...560L..95C}
{Chae}, J. 2001, \apjl, 560, L95, \dodoi{10.1086/324173}

\bibitem[{{Chintzoglou} \& {Zhang}(2013)}]{2013ApJ...764L...3C}
{Chintzoglou}, G., \& {Zhang}, J. 2013, \apjl, 764, L3,
  \dodoi{10.1088/2041-8205/764/1/L3}

\bibitem[{{D{\'e}moulin} \& {Pariat}(2009)}]{2009AdSpR..43.1013D}
{D{\'e}moulin}, P., \& {Pariat}, E. 2009, Advances in Space Research, 43, 1013,
  \dodoi{10.1016/j.asr.2008.12.004}

\bibitem[{{Fang} \& {Fan}(2015)}]{2015ApJ...806...79F}
{Fang}, F., \& {Fan}, Y. 2015, \apj, 806, 79,
  \dodoi{10.1088/0004-637X/806/1/79}

\bibitem[{{Finn} \& {Antonsen}(1985)}]{1985CoPPC..9..111}
{Finn}, J.~M., \& {Antonsen}, T.~M.~J. 1985, Comments Plasma Phys. Controlled
  Fusion; (United Kingdom), 9, 111

\bibitem[{{Gaizauskas} {et~al.}(1983){Gaizauskas}, {Harvey}, {Harvey}, \&
  {Zwaan}}]{1983ApJ...265.1056G}
{Gaizauskas}, V., {Harvey}, K.~L., {Harvey}, J.~W., \& {Zwaan}, C. 1983, \apj,
  265, 1056, \dodoi{10.1086/160747}

\bibitem[{{Georgoulis} \& {Rust}(2007)}]{2007ApJ...661L.109G}
{Georgoulis}, M.~K., \& {Rust}, D.~M. 2007, \apjl, 661, L109,
  \dodoi{10.1086/518718}

\bibitem[{{Gyenge} {et~al.}(2016){Gyenge}, {Ludm{\'a}ny}, \&
  {Baranyi}}]{2016ApJ...818..127G}
{Gyenge}, N., {Ludm{\'a}ny}, A., \& {Baranyi}, T. 2016, \apj, 818, 127,
  \dodoi{10.3847/0004-637X/818/2/127}

\bibitem[{{Gyenge} {et~al.}(2017){Gyenge}, {Singh}, {Kiss}, {Srivastava}, \&
  {Erd{\'e}lyi}}]{2017ApJ...838...18G}
{Gyenge}, N., {Singh}, T., {Kiss}, T.~S., {Srivastava}, A.~K., \&
  {Erd{\'e}lyi}, R. 2017, \apj, 838, 18, \dodoi{10.3847/1538-4357/aa62a8}

\bibitem[{{Hagino} \& {Sakurai}(2005)}]{2005PASJ...57..481H}
{Hagino}, M., \& {Sakurai}, T. 2005, \pasj, 57, 481,
  \dodoi{10.1093/pasj/57.3.481}

\bibitem[{{Hoeksema} {et~al.}(2014){Hoeksema}, {Liu}, {Hayashi}, {Sun},
  {Schou}, {Couvidat}, {Norton}, {Bobra}, {Centeno}, {Leka}, {Barnes}, \&
  {Turmon}}]{2014SoPh..289.3483H}
{Hoeksema}, J.~T., {Liu}, Y., {Hayashi}, K., {et~al.} 2014, \solphys, 289,
  3483, \dodoi{10.1007/s11207-014-0516-8}

\bibitem[{{Hotta} \& {Toriumi}(2020)}]{2020MNRAS.498.2925H}
{Hotta}, H., \& {Toriumi}, S. 2020, \mnras, 498, 2925,
  \dodoi{10.1093/mnras/staa2529}

\bibitem[{{Joshi} \& {Chandra}(2019)}]{2019OAst...28..228J}
{Joshi}, A., \& {Chandra}, R. 2019, Open Astronomy, 28, 228,
  \dodoi{10.1515/astro-2019-0019}

\bibitem[{{Komm} \& {Gosain}(2015)}]{2015ApJ...798...20K}
{Komm}, R., \& {Gosain}, S. 2015, \apj, 798, 20,
  \dodoi{10.1088/0004-637X/798/1/20}

\bibitem[{{Komm} \& {Gosain}(2019)}]{2019ApJ...887..192K}
---. 2019, \apj, 887, 192, \dodoi{10.3847/1538-4357/ab58ca}

\bibitem[{{Kusano} {et~al.}(2020){Kusano}, {Iju}, {Bamba}, \&
  {Inoue}}]{2020Sci...369..587K}
{Kusano}, K., {Iju}, T., {Bamba}, Y., \& {Inoue}, S. 2020, Science, 369, 587,
  \dodoi{10.1126/science.aaz2511}

\bibitem[{{LaBonte} {et~al.}(2007){LaBonte}, {Georgoulis}, \&
  {Rust}}]{2007ApJ...671..955L}
{LaBonte}, B.~J., {Georgoulis}, M.~K., \& {Rust}, D.~M. 2007, \apj, 671, 955,
  \dodoi{10.1086/522682}

\bibitem[{{Lee} {et~al.}(2018){Lee}, {Park}, \& {Moon}}]{2018SoPh..293..159L}
{Lee}, E.-J., {Park}, S.-H., \& {Moon}, Y.-J. 2018, \solphys, 293, 159,
  \dodoi{10.1007/s11207-018-1381-7}

\bibitem[{{Lee} {et~al.}(2012){Lee}, {Moon}, {Lee}, {Lee}, \&
  {Na}}]{2012SoPh..281..639L}
{Lee}, K., {Moon}, Y.~J., {Lee}, J.-Y., {Lee}, K.-S., \& {Na}, H. 2012,
  \solphys, 281, 639, \dodoi{10.1007/s11207-012-0091-9}

\bibitem[{{Leka} \& {Barnes}(2003)}]{2003ApJ...595.1296L}
{Leka}, K.~D., \& {Barnes}, G. 2003, \apj, 595, 1296, \dodoi{10.1086/377512}

\bibitem[{{Leka} \& {Barnes}(2007)}]{2007ApJ...656.1173L}
---. 2007, \apj, 656, 1173, \dodoi{10.1086/510282}

\bibitem[{{Leka} {et~al.}(1996){Leka}, {Canfield}, {McClymont}, \& {van
  Driel-Gesztelyi}}]{1996ApJ...462..547L}
{Leka}, K.~D., {Canfield}, R.~C., {McClymont}, A.~N., \& {van Driel-Gesztelyi},
  L. 1996, \apj, 462, 547, \dodoi{10.1086/177171}

\bibitem[{{Leka} {et~al.}(2019){Leka}, {Park}, {Kusano}, {Andries}, {Barnes},
  {Bingham}, {Bloomfield}, {McCloskey}, {Delouille}, {Falconer}, {Gallagher},
  {Georgoulis}, {Kubo}, {Lee}, {Lee}, {Lobzin}, {Mun}, {Murray}, {Hamad
  Nageem}, {Qahwaji}, {Sharpe}, {Steenburgh}, {Steward}, \&
  {Terkildsen}}]{2019ApJS..243...36L}
{Leka}, K.~D., {Park}, S.-H., {Kusano}, K., {et~al.} 2019, \apjs, 243, 36,
  \dodoi{10.3847/1538-4365/ab2e12}

\bibitem[{{Li} {et~al.}(2019){Li}, {Xiang}, {Xie}, \&
  {Xu}}]{2019ApJ...873..121L}
{Li}, F.~Y., {Xiang}, N.~B., {Xie}, J.~L., \& {Xu}, J.~C. 2019, \apj, 873, 121,
  \dodoi{10.3847/1538-4357/ab06bf}

\bibitem[{{Li} {et~al.}(2002){Li}, {Wang}, {Xiong}, {Liang}, {Yun}, \&
  {Gu}}]{2002A&A...383..648L}
{Li}, K.~J., {Wang}, J.~X., {Xiong}, S.~Y., {et~al.} 2002, \aap, 383, 648,
  \dodoi{10.1051/0004-6361:20011799}

\bibitem[{{Liu} {et~al.}(2017){Liu}, {Deng}, {Wang}, \&
  {Wang}}]{2017ApJ...843..104L}
{Liu}, C., {Deng}, N., {Wang}, J. T.~L., \& {Wang}, H. 2017, \apj, 843, 104,
  \dodoi{10.3847/1538-4357/aa789b}

\bibitem[{{Liu} {et~al.}(2014){Liu}, {Hoeksema}, \&
  {Sun}}]{2014ApJ...783L...1L}
{Liu}, Y., {Hoeksema}, J.~T., \& {Sun}, X. 2014, \apjl, 783, L1,
  \dodoi{10.1088/2041-8205/783/1/L1}

\bibitem[{{Livi} {et~al.}(1989){Livi}, {Martin}, {Wang}, \&
  {Ai}}]{1989SoPh..121..197L}
{Livi}, S. H.~B., {Martin}, S., {Wang}, H., \& {Ai}, G. 1989, \solphys, 121,
  197, \dodoi{10.1007/BF00161696}

\bibitem[{{Longcope} {et~al.}(1998){Longcope}, {Fisher}, \&
  {Pevtsov}}]{1998ApJ...507..417L}
{Longcope}, D.~W., {Fisher}, G.~H., \& {Pevtsov}, A.~A. 1998, \apj, 507, 417,
  \dodoi{10.1086/306312}

\bibitem[{{Longcope} {et~al.}(2007){Longcope}, {Ravindra}, \&
  {Barnes}}]{2007ApJ...668..571L}
{Longcope}, D.~W., {Ravindra}, B., \& {Barnes}, G. 2007, \apj, 668, 571,
  \dodoi{10.1086/521095}

\bibitem[{{Mandal} {et~al.}(2017){Mandal}, {Chatterjee}, \&
  {Banerjee}}]{2017ApJ...835...62M}
{Mandal}, S., {Chatterjee}, S., \& {Banerjee}, D. 2017, \apj, 835, 62,
  \dodoi{10.3847/1538-4357/835/1/62}

\bibitem[{{Nishizuka} {et~al.}(2017){Nishizuka}, {Sugiura}, {Kubo}, {Den},
  {Watari}, \& {Ishii}}]{2017ApJ...835..156N}
{Nishizuka}, N., {Sugiura}, K., {Kubo}, Y., {et~al.} 2017, \apj, 835, 156,
  \dodoi{10.3847/1538-4357/835/2/156}

\bibitem[{{Park} {et~al.}(2010){Park}, {Chae}, \& {Wang}}]{2010ApJ...718...43P}
{Park}, S.-H., {Chae}, J., \& {Wang}, H. 2010, \apj, 718, 43,
  \dodoi{10.1088/0004-637X/718/1/43}

\bibitem[{{Park} {et~al.}(2008){Park}, {Lee}, {Choe}, {Chae}, {Jeong}, {Yang},
  {Jing}, \& {Wang}}]{2008ApJ...686.1397P}
{Park}, S.-H., {Lee}, J., {Choe}, G.~S., {et~al.} 2008, \apj, 686, 1397,
  \dodoi{10.1086/591117}

\bibitem[{{Park} {et~al.}(2020){Park}, {Leka}, \&
  {Kusano}}]{2020ApJ...904....6P}
{Park}, S.-H., {Leka}, K.~D., \& {Kusano}, K. 2020, \apj, 904, 6,
  \dodoi{10.3847/1538-4357/abbb93}

\bibitem[{{Pesnell} {et~al.}(2012){Pesnell}, {Thompson}, \&
  {Chamberlin}}]{2012SoPh..275....3P}
{Pesnell}, W.~D., {Thompson}, B.~J., \& {Chamberlin}, P.~C. 2012, \solphys,
  275, 3, \dodoi{10.1007/s11207-011-9841-3}

\bibitem[{{Pevtsov} {et~al.}(2001){Pevtsov}, {Canfield}, \&
  {Latushko}}]{2001ApJ...549L.261P}
{Pevtsov}, A.~A., {Canfield}, R.~C., \& {Latushko}, S.~M. 2001, \apjl, 549,
  L261, \dodoi{10.1086/319179}

\bibitem[{{Romano} {et~al.}(2014){Romano}, {Zuccarello}, {Guglielmino}, \&
  {Zuccarello}}]{2014ApJ...794..118R}
{Romano}, P., {Zuccarello}, F.~P., {Guglielmino}, S.~L., \& {Zuccarello}, F.
  2014, \apj, 794, 118, \dodoi{10.1088/0004-637X/794/2/118}

\bibitem[{{Sammis} {et~al.}(2000){Sammis}, {Tang}, \&
  {Zirin}}]{2000ApJ...540..583S}
{Sammis}, I., {Tang}, F., \& {Zirin}, H. 2000, \apj, 540, 583,
  \dodoi{10.1086/309303}

\bibitem[{{Scherrer} {et~al.}(2012){Scherrer}, {Schou}, {Bush}, {Kosovichev},
  {Bogart}, {Hoeksema}, {Liu}, {Duvall}, {Zhao}, {Title}, {Schrijver},
  {Tarbell}, \& {Tomczyk}}]{2012SoPh..275..207S}
{Scherrer}, P.~H., {Schou}, J., {Bush}, R.~I., {et~al.} 2012, \solphys, 275,
  207, \dodoi{10.1007/s11207-011-9834-2}

\bibitem[{{Schuck}(2008)}]{2008ApJ...683.1134S}
{Schuck}, P.~W. 2008, \apj, 683, 1134, \dodoi{10.1086/589434}

\bibitem[{{Toriumi} \& {Hotta}(2019)}]{2019ApJ...886L..21T}
{Toriumi}, S., \& {Hotta}, H. 2019, \apjl, 886, L21,
  \dodoi{10.3847/2041-8213/ab55e7}

\bibitem[{{Toriumi} {et~al.}(2014){Toriumi}, {Iida}, {Kusano}, {Bamba}, \&
  {Imada}}]{2014SoPh..289.3351T}
{Toriumi}, S., {Iida}, Y., {Kusano}, K., {Bamba}, Y., \& {Imada}, S. 2014,
  \solphys, 289, 3351, \dodoi{10.1007/s11207-014-0502-1}

\bibitem[{{Toriumi} \& {Takasao}(2017)}]{2017ApJ...850...39T}
{Toriumi}, S., \& {Takasao}, S. 2017, \apj, 850, 39,
  \dodoi{10.3847/1538-4357/aa95c2}

\bibitem[{{Toriumi} \& {Wang}(2019)}]{2019LRSP...16....3T}
{Toriumi}, S., \& {Wang}, H. 2019, Living Reviews in Solar Physics, 16, 3,
  \dodoi{10.1007/s41116-019-0019-7}

\bibitem[{{Vemareddy} \& {D{\'e}moulin}(2017)}]{2017A&A...597A.104V}
{Vemareddy}, P., \& {D{\'e}moulin}, P. 2017, \aap, 597, A104,
  \dodoi{10.1051/0004-6361/201629282}

\bibitem[{{Zhang}(2006)}]{2006ApJ...646L..85Z}
{Zhang}, M. 2006, \apjl, 646, L85, \dodoi{10.1086/506560}

\bibitem[{{Zhongxian} \& {Jingxiu}(1994)}]{1994SoPh..149..105S}
{Zhongxian}, S., \& {Jingxiu}, W. 1994, \solphys, 149, 105,
  \dodoi{10.1007/BF00645181}

\bibitem[{{Zirin} \& {Liggett}(1987)}]{1987SoPh..113..267Z}
{Zirin}, H., \& {Liggett}, M.~A. 1987, \solphys, 113, 267,
  \dodoi{10.1007/BF00147707}

\end{thebibliography}
\end{document}